\documentstyle[prb,aps,twocolumn]{revtex}
\input epsf
\scrollmode
\begin{document}
\twocolumn[\hsize\textwidth\columnwidth\hsize\csname
@twocolumnfalse\endcsname

\title{Comment on ``Quantum Monte Carlo study of the dipole moment of
CO'' [J. Chem. Phys. 110, 11700 (1999)]}

\author{K.~C.~Huang\cite{address}, R.~J.~Needs, and G.~Rajagopal}

\address{Cavendish Laboratory, University of Cambridge, Madingley
Road, Cambridge CB3 0HE, U.K.}

\date{\today}

\maketitle

\begin{abstract}
\begin{quote}
\parbox{16 cm} {\small }
\end{quote}
\end{abstract}

\pacs{PACS: }
]

\narrowtext

In a recent paper Schautz and Flad~\cite{schautz_1999} studied the
dipole moment of CO using various first-principles computational
methods.  As part of this study they considered the applicability of
the Hellmann-Feynman theorem within the fixed-node diffusion quantum
Monte Carlo (DMC) method~\cite{dmc,hammond,reynolds_1982}.  The
purpose of this comment is to point out that the conclusion reached by
Schautz and Flad~\cite{schautz_1999} regarding the applicability of
the Hellmann-Feynman theorem within fixed-node DMC is incorrect.

In the DMC method stochastic evolution of the imaginary-time
Schr\"{o}dinger equation is used to project out the lowest energy
many-electron wave function.  Such projector methods suffer from a
fermion sign problem in which the wave function decays towards the
bosonic ground state.  To enforce the fermion symmetry the fixed-node
approximation~\cite{anderson_1975,anderson_1976} is normally used.  In
a fixed-node DMC calculation the Schr\"{o}dinger equation is solved
separately in each nodal pocket with the boundary condition that the
wave function in the $\alpha^{th}$ pocket, $\psi_{\alpha}$, is zero on
the surface of the pocket.  For simplicity we consider only ground
state wave functions which satisfy the tiling
property~\cite{ceperley_1991,foulkes_1999} that all nodal pockets are
equivalent and related by the permutation symmetry.  The pocket wave
function is zero outside the pocket and has a discontinuous derivative
at the surface of the pocket and therefore
satisfies~\cite{foulkes_1999}

\begin{equation}
\label{SE_pocket}
\hat{H} \psi_{\alpha} = E_{\rm D} \psi_{\alpha} +
g[\psi_{\alpha}]\,\delta({\bf r} - {\bf r}_{\rm n}[\psi_{\alpha}]) \;,
\end{equation}

\noindent where ${\bf r}_{\rm n}[\psi_{\alpha}]$ is the surface of
pocket $\alpha$.  The delta function term, which was neglected by
Schautz and Flad~\cite{schautz_1999}, arises from the action of the
kinetic energy operator on the discontinuity in the derivative of the
wave function at the pocket surface.  Operating on this equation with
the antisymmetrising operator, $\hat{A}$, gives

\begin{equation}
\label{SE_anti}
\hat{H} \Psi = E_{\rm D} \Psi + h[\Psi]\,\delta({\bf r} - {\bf r}_{\rm
n}[\Psi]) \;,
\end{equation}

\noindent where $\Psi = \hat{A}\psi_{\alpha}$ is the antisymmetric DMC
wave function and ${\bf r}_{\rm n}[\Psi]$ is the nodal surface of
$\Psi$.  The nodal surface is fixed by the choice of the trial wave
function, $\Phi_T$.  The DMC energy may then be calculated from

\begin{equation}
\label{E_D}
E_{\rm D} = \frac{\int \Phi_T \hat{H} \Psi \, d{\bf r}}{\int
\Phi_T\Psi \, d{\bf r}}\;.
\end{equation}

If the nodal surface of $\Phi_T$ is exact then $\Psi$ has no gradient
discontinuities on the nodal surface and $\hat{H} \Psi_0 = E_0
\Psi_0$, where $\Psi_0$ and $E_0$ are the exact wave function and
energy, but if the nodal surface is inexact then $h[\Psi]$ is non-zero
and normally of order $\Psi-\Psi_0$.  On substituting
Eq.~\ref{SE_anti} into Eq.~\ref{E_D} we find that the nodal term,
$h\delta$, does not contribute to the energy because $\Phi_T$ is zero
on the nodal surface.  However, the nodal term {\it can} contribute to
derivatives of $E_{\rm D}$.  A simple way to see this is to note that
the DMC energy can also be evaluated as the expectation value with the
fixed-node DMC wave function,

\begin{equation}
\label{E_D2}
E_{\rm D} = \frac{\int \Psi\hat{H}\Psi \, d{\bf r}}{\int \Psi\Psi \,
d{\bf r}}\;.
\end{equation}

\noindent The derivative with respect to a parameter, $\lambda$, is

\begin{equation}
\label{dED/dR_nodes1}
\frac{\partial E_{\rm D}}{\partial\lambda} = \frac{\int \Psi
\frac{\partial\hat{H}}{\partial\lambda} \Psi \, d{\bf r}} {\int
\Psi\Psi \, d{\bf r}} + 2 \frac{\int
\frac{\partial\Psi}{\partial\lambda} \hat{H} \Psi \, d{\bf r}} {\int
\Psi\Psi \, d{\bf r}} - 2 E_{\rm D} \frac{\int \Psi
\frac{\partial\Psi}{\partial\lambda} \, d{\bf r}} {\int \Psi\Psi \,
d{\bf r}} \;,
\end{equation}

\noindent and using Eq.~\ref{SE_anti} we obtain

\begin{equation}
\label{dED/dR_nodes2}
\frac{\partial E_{\rm D}} {\partial\lambda} = \frac{\int \Psi
\frac{\partial\hat{H}}{\partial\lambda} \Psi \, d{\bf r}} {\int
\Psi\Psi \, d{\bf r}} + 2 \frac{\int
\frac{\partial\Psi}{\partial\lambda} h\delta \, d{\bf r}} {\int
\Psi\Psi \, d{\bf r}} \;,
\end{equation}

\noindent which is the Hellmann-Feynman expression with an additional
nodal term.  While $h\delta$ is determined by the nodes of $\Psi$ (or
equivalently those of $\Phi_T$),
$\frac{\partial\Psi}{\partial\lambda}$ depends on how we choose the
nodes to vary with $\lambda$.  We therefore expect the nodal term in
Eq.~\ref{dED/dR_nodes2} to be non-zero in general.  If we choose the
nodes to be independent of $\lambda$ then
$\frac{\partial\Psi}{\partial\lambda}=0$ on the nodal surface and the
contribution from the nodal term is zero.  In this case the
Hellmann-Feynman theorem holds, as correctly stated by Schautz and
Flad~\cite{schautz_1999}.  However, if the nodes of the trial wave
function vary with $\lambda$ then the nodal term will normally be
non-zero and the Hellmann-Feynman expression does not give the exact
derivative of the DMC energy of Eqs.~\ref{E_D} and~\ref{E_D2}, in
contradiction to Schautz and Flad~\cite{schautz_1999}.

We conclude that Schautz and Flad are correct in stating that the
Hellmann-Feynman expression evaluated with the fixed-node DMC wave
function is equal to the derivative of the fixed-node DMC energy with
respect to a parameter $\lambda$ {\it if the nodes are independent of
$\lambda$}, but this does not hold in general if the nodal surface
depends on $\lambda$.

\end{document}